\documentclass[twocolumn]{aastex62}
\pdfoutput=1 
\usepackage{amsmath,amstext}
\usepackage[T1]{fontenc}
\usepackage[figure,figure*]{hypcap}
\usepackage[version=4]{mhchem} 

\turnoffedits

\newcommand{\rosetta}{{\sl Rosetta}}

\graphicspath{{./}{figures/}}

\shorttitle{\rosetta\ Upper Limits}
\shortauthors{Keeney et~al.}

\begin{document}

\title{Upper Limits for Emissions in the Coma of Comet 67P/Churyumov-Gerasimenko Near Perihelion as Measured by \rosetta's Alice Far-Ultraviolet Spectrograph}

\author[0000-0003-0797-5313]{Brian A. Keeney}
\affiliation{Southwest Research Institute, Department of Space Studies, Suite 300, 1050 Walnut St., Boulder, CO 80302, USA}
\email{bkeeney@gmail.com}
\author{S. Alan Stern}
\affiliation{Southwest Research Institute, Department of Space Studies, Suite 300, 1050 Walnut St., Boulder, CO 80302, USA}
\author{Ronald J. Vervack, Jr.}
\affiliation{Space Exploration Sector, Johns Hopkins University Applied Physics Laboratory, 11100 Johns Hopkins Rd., Laurel, MD 20723, USA}
\author{Matthew M. Knight}
\affiliation{Department of Astronomy, University of Maryland, College Park, MD 20742, USA}
\author{John Noonan}
\affiliation{Lunar and Planetary Laboratory, University of Arizona, 1629 E. University Blvd., Tucson, AZ 85721, USA}
\author{Joel Wm. Parker}
\affiliation{Southwest Research Institute, Department of Space Studies, Suite 300, 1050 Walnut St., Boulder, CO 80302, USA}
\author{Michael F. A'Hearn}
\altaffiliation{Deceased}
\affiliation{Department of Astronomy, University of Maryland, College Park, MD 20742, USA}
\author{Jean-Loup Bertaux}
\affiliation{LATMOS, CNRS/UVSQ/IPSL, 11 Boulevard d'Alembert, F-78280 Guyancourt, France}
\author{Lori M. Feaga}
\affiliation{Department of Astronomy, University of Maryland, College Park, MD 20742, USA}
\author{Paul D. Feldman}
\affiliation{Department of Physics and Astronomy, Johns Hopkins University, 3400 N. Charles St., Baltimore, MD 21218, USA}
\author{Richard A. Medina}
\affiliation{Southwest Research Institute, Department of Space Studies, Suite 300, 1050 Walnut St., Boulder, CO 80302, USA}
\author{Jon P. Pineau}
\affiliation{Stellar Solutions, Inc., 250 Cambridge Ave., Suite 204, Palo Alto, CA 94306, USA}
\author{Rebecca N. Schindhelm}
\affiliation{Southwest Research Institute, Department of Space Studies, Suite 300, 1050 Walnut St., Boulder, CO 80302, USA}
\affiliation{Ball Aerospace and Technology Corp., 1600 Commerce St., Boulder, CO 80301, USA}
\author{Andrew J. Steffl}
\affiliation{Southwest Research Institute, Department of Space Studies, Suite 300, 1050 Walnut St., Boulder, CO 80302, USA}
\author{M. Versteeg}
\affiliation{Southwest Research Institute, 6220 Culebra Rd., San Antonio, TX 78238, USA}
\author{Harold A. Weaver}
\affiliation{Space Exploration Sector, Johns Hopkins University Applied Physics Laboratory, 11100 Johns Hopkins Rd., Laurel, MD 20723, USA}

\begin{abstract}
The Alice far-UV imaging spectrograph (700-2050~\AA) acquired over 70,000 spectral images during \rosetta's 2-year escort mission, including over 20,000 in the months surrounding perihelion when the comet activity level was highest. We have developed automated software to fit and remove ubiquitous H, O, C, S, and CO emissions from Alice spectra, along with reflected solar continuum and absorption from gaseous \ce{H2O} in the comet's coma, which we apply to a ``grand sum'' of integrations taken near perihelion. We present upper limits on the presence of one ion and 17~neutral atomic species for this time period. These limits are compared to results obtained by other \rosetta\ instruments where possible, as well as to CI carbonaceous chondrites and solar photospheric abundances.
\end{abstract}

\keywords{comets: individual (67P) -- ultraviolet: planetary systems}

\section{Introduction}
\label{sec:intro}

A stunning variety of volatile species have been found in the coma of Comet 67P/Churyumov-Gerasimenko (67P/C-G) by the instruments on the \rosetta\ orbiter \citep{bieler15,leroy15,altwegg17,calmonte17,hassig17}. Although the initial discoveries have been made by the \rosetta\ Orbiter Spectrometer for Ion and Neutral Analysis' Double-Focusing Mass Spectrograph \citep[ROSINA/DFMS;][]{balsiger07}, remote-sensing instruments like the Alice ultraviolet spectrograph \citep{stern07}, the Visible and Infrared Thermal Imaging Spectrometer \citep[VIRTIS;][]{coradini07}, and the Microwave Instrument for the \rosetta\ Orbiter \citep[MIRO;][]{gulkis07} have provided important and independent confirmations \citep{biver15,bockelee-morvan15,bockelee-morvan16,feldman15,migliorini16,keeney17,marshall17}.

An essential part of any inventory is setting limits on species that could have been detected but were not. In this regard, Alice is well-positioned among \rosetta's remote-sensing instruments due to its far-UV bandpass \citep[700-2050~\AA;][]{stern07} that includes the strongest resonance lines of many neutral and singly-ionized atoms. This sensitivity is double-edged, however, because in order to search for signals from weak, undetected species, ubiquitous emissions from the daughter products of the primary parent molecules \citep[\ce{H2O}, \ce{CO2}, \ce{CO}, and \ce{O2};][]{fougere16} must first be modeled and removed (e.g., H 1026, 1216~\AA; O 1304, 1356~\AA; C 1561, 1657~\AA). Further complicating matters, CO Fourth Positive (1300-1700~\AA) and Cameron (1900-2100~\AA) band emissions are also prevalent in Alice data near perihelion \citep{feldman16,feldman18}.

After these ubiquitous emissions are removed, we calculate upper limits on 18~species in the coma of 67P/C-G near perihelion. \autoref{sec:obs} details our observations. \autoref{sec:sum} presents the high-S/N ``grand sum'' used to constrain the presence of undetected species, and our method to fit and remove detected emissions. \autoref{sec:limits} presents our upper limits. \autoref{sec:disc} compares our results with previous measurements, and \autoref{sec:summary} summarizes our conclusions.

\section{Observations}
\label{sec:obs}

We focus our analysis on the timeframe near perihelion where the comet's activity level was highest \citep{fougere16,hansen16,lauter19} so we can set the most sensitive limits on undetected species with respect to \ce{H2O}, the primary parent molecule. In all, Alice obtained 20,300~spectral images within $\pm90$~days of perihelion, with a total exposure time, $t_{\rm exp}$, of 98.8~days. During this time, the median and average exposure times were 5 and 7~minutes, respectively, but ranged from 0.1-60~minutes. However, some of these exposures did not have stable pointing or were calibration exposures that did not contain the nucleus \citep{pineau19} and are not useful for our purposes. Removing these exposures yields a dataset of 12,400~spectral images with $t_{\rm exp}=64.0$~days.

\begin{figure}
  \fig{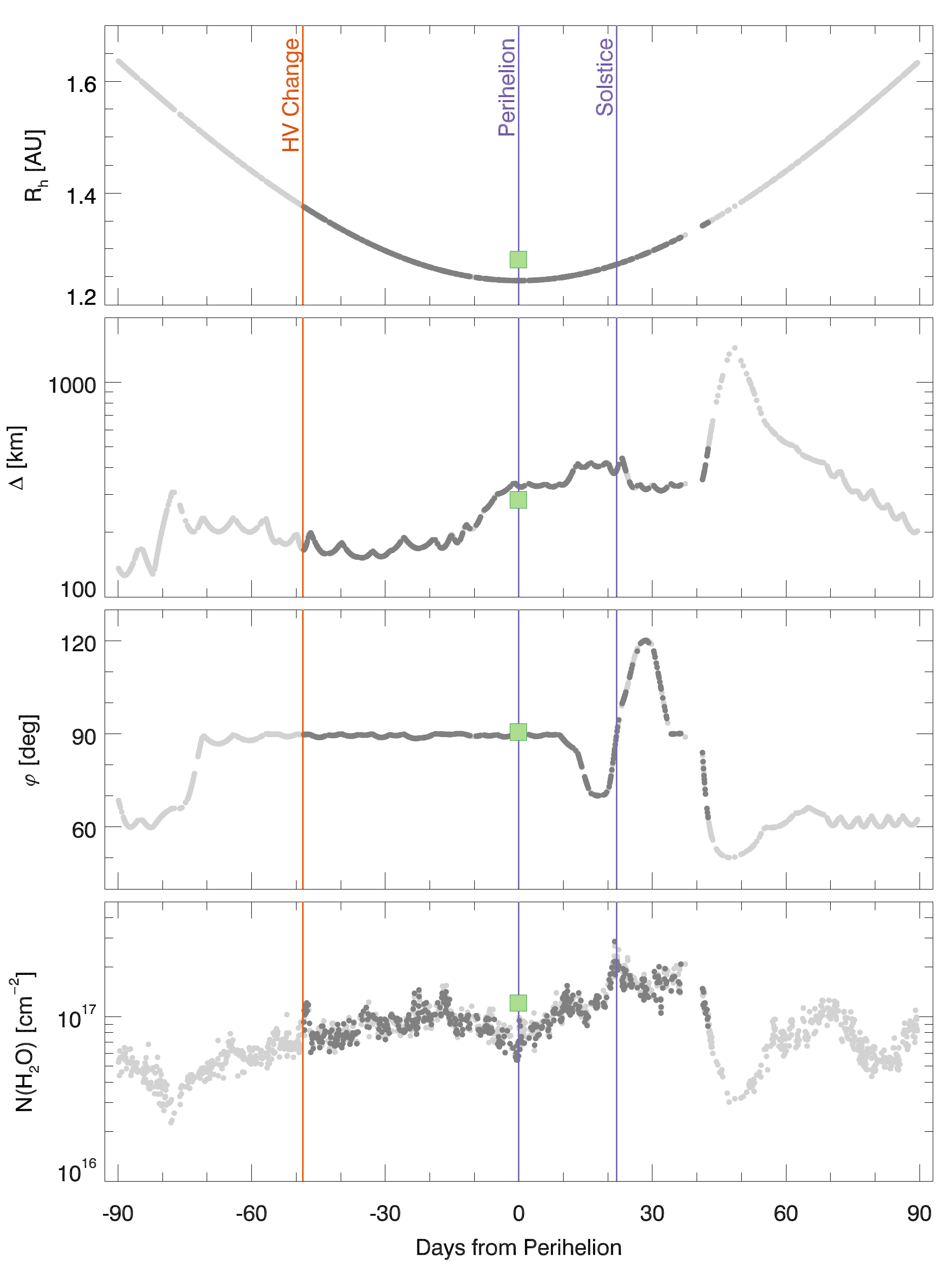}{\columnwidth}{}
  \vspace{-3em}
  \caption{Variation in observing geometry near 67P/C-G's perihelion passage of 2015 August~13. From top to bottom, the panels show the heliocentric distance, $R_h$, the spacecraft-comet distance, $\Delta$, the solar phase angle, $\phi$, and the column density of water, $N_{\ce{H2O}}$, as a function of time. The gray points show exposures that include the nucleus and have stable pointing, with the dark gray points showing the subset of exposures that contribute to our ``grand sum'' (\autoref{sec:sum}); the large green squares show the weighted mean values for the grand sum, where each exposure is weighted by $N_{\ce{H2O}}\,t_{\rm exp}^{1/2}/R_h^2$. The bottom panel uses a modified version of the \citet{hansen16} empirical coma model to predict $N_{\ce{H2O}}$ at a given slit position in Alice exposures (see text for details). The locations of perihelion, solstice, and an Alice high-voltage change \citep{pineau19} are also shown.
    \label{fig:geometry}}
\end{figure}

\autoref{fig:geometry} displays the observing geometry near perihelion. There is little variation during most of this time, particularly in phase angle, $\phi$,  which was $\approx90\degr$ from early June ($\sim70$~days pre-perihelion) until late August ($\sim10$~days post-perihelion) when the spacecraft was executing so-called ``terminator'' orbits \citep{pineau19}. The spacecraft-comet distance, $\Delta$, ranges from $\Delta\sim150$-500~km until $\sim40$~days after perihelion, when \rosetta\ began a tail excursion. Just before the tail excursion there is a gap from 2015 September~19-22 (27-30~days post-perihelion), where none of the \rosetta\ instruments collected data due to a spacecraft maintenance issue.

\section{Creation of Grand Sum}
\label{sec:sum}

\begin{figure}
  \fig{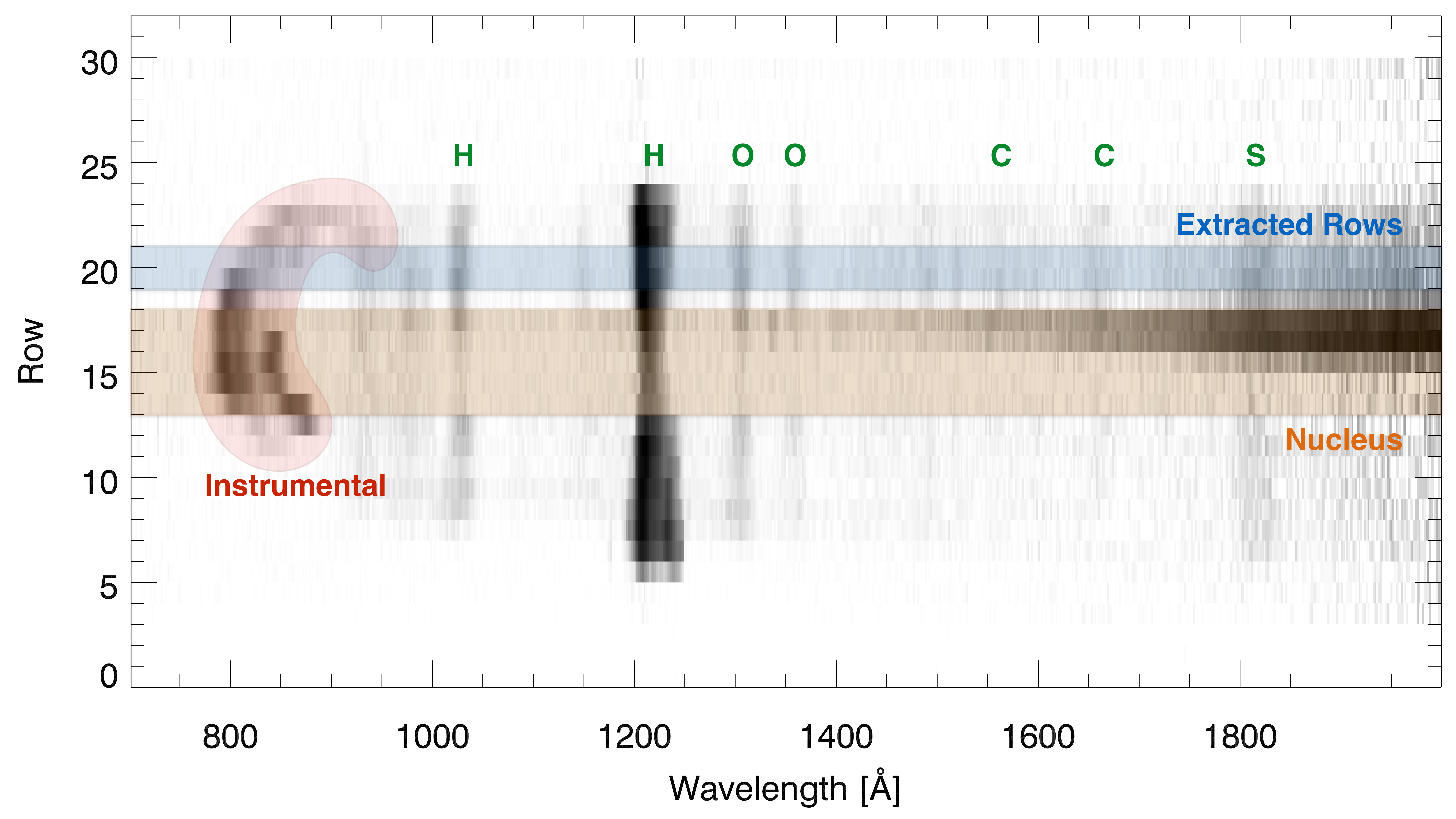}{\columnwidth}{}
  \vspace{-3em}
  \caption{An example Alice spectral image acquired at 11:22:51~UT on 2015 July~1, when 67P/C-G was 1.35~AU from the Sun, \rosetta\ was 160~km from the comet center, and the solar phase angle was $89\fdg6$. The nucleus (orange shaded region) and extracted coma rows (blue shaded region) are labeled, along with the positions of several atomic emission lines and the instrumental feature (red shaded region) affecting wavelengths $\lesssim1000$~\AA\ \citep{noonan16}.
  \label{fig:spec2d}}
\end{figure}

\autoref{fig:spec2d} shows an Alice spectral image with the nucleus, several common emission lines, and a strong instrumental feature \citep{noonan16} identified. The Alice slit is $5\fdg5$ long and has a dog-bone shape that is twice as wide at the top and bottom as in the center \citep{stern07}, which is evident in the shape of the emission lines, particularly H Ly$\alpha$ at 1216~\AA.

For each exposure in the timeframe of interest, we identify the location of the nucleus, which is brightest in long-wavelength reflected solar light. We then extract the two coma rows closest to the sunward\footnote{The sunward direction is toward the top (i.e., higher row numbers) in Alice spectral images.} edge of the nucleus, padding by one row from the identified edge to ensure that we are extracting only coma emissions. These steps are performed automatically using SPICE kernel information propagated to the individual FITS headers, before being vetted and adjusted manually. This vetting is essential due to contamination from UV-bright stars occasionally entering the Alice slit, instances where the instrumental artifact is unusually bright and/or extended, and cases where the position of the nucleus on the Alice slit varied during the exposure.

The bottom panel of \autoref{fig:geometry} shows an estimate of the \ce{H2O} column density for the extracted rows in each Alice exposure using a slightly modified version of the empirical coma model of \citet{hansen16}. We showed in \citet{keeney19} that the column density of water, $N_{\ce{H2O}}$, measured by Alice in absorption against the continuum of UV-bright stars passing near the nucleus of 67P/C-G is consistent with the predictions of the \citet{hansen16} model; however, we noted that our measurements disagreed with their prediction very close to perihelion where their parameterization of the \ce{H2O} production rate, $Q_{\ce{H2O}}$, is discontinuous. Here we remove this discontinuity by adopting the \citet{hansen16} parameterization of $Q_{\ce{H2O}}$ before perihelion and after solstice, and linearly interpolating in-between. This modification yields model column densities that more closely agree with the Alice measurements \citep{keeney17,keeney19} and has a maximum $Q_{\ce{H2O}} = 2.9\times10^{28}~\mathrm{s^{-1}}$ occurring 22~days after perihelion, which is consistent with the peak values measured by ROSINA \citep[$Q_{\ce{H2O}} = 3.5\pm0.5 \times 10^{28}~\mathrm{s^{-1}}$ occurring 18-22~days after perihelion;][]{hansen16}. \edit1{Note that despite the agreement between the Alice measurements and the \citet{hansen16} model, there is not yet a consensus on $Q_{\ce{H2O}}$ from \rosetta\ measurements.}

After extracting coma rows from each of the vetted Alice exposures, we must decide which exposures should contribute to our ``grand sum''. Our goal is to create a coaddition that yields the most stringent limits for each species compared to \ce{H2O}, which requires maximizing three competing factors: (1) the S/N ratio of the resulting sum, which is proportional to the square root of the total exposure time; (2) the fluorescence efficiency ($g$) factors, which are maximized at perihelion when the intensity of solar radiation is highest; and (3) $N_{\ce{H2O}}$, which is largest at solstice (\autoref{fig:geometry}). Strictly maximizing one of these factors is detrimental to the other two (e.g., maximizing S/N by including as many exposures as possible will dilute the average $g$-factors and $N_{\ce{H2O}}$ of the sum), so a compromise solution must be found.

When seeking this compromise, there are two other factors that must be considered. The first is that the typical location of the nucleus in the Alice slit changes during this time period, with the nucleus more likely to be located near the bottom of the slit at earlier times and near the top of the slit at later times. Because the spectral resolution of the top and bottom regions of the slit is worse than the resolution in the narrower central region \citep[see \autoref{fig:spec2d};][]{stern07} and we wish to minimize systematic effects introduced by combining spectra with different resolutions, this consideration causes us to prefer exposures taken in the middle of the time frame. The second factor is that Alice changed its high-voltage setting on 2015 June~25 to mitigate the effects of gain sag at the previous setting \citep{pineau19}; thus, we avoid exposures taken prior to this date.

Given these constraints, we chose the timeframe from 48.5 days before perihelion (when the high-voltage setting changed) to 42.5 days after perihelion (when \rosetta\ began its tail excursion) to create our sum. During this time, the heliocentric distance, $R_h$, ranged from 1.24 to 1.38~AU, $\Delta$ ranged from 150 to 490~km, $\phi$ ranged from 63 to $120\degr$, and $N_{\ce{H2O}}$ ranged from $3.6\times10^{16}$ to $3.1\times10^{17}~\mathrm{cm}^{-2}$ (\autoref{fig:geometry}). In addition, we only consider exposures where the extracted rows are in the range 12-19 to minimize blending of different spectral resolutions while including sufficient exposures to produce a high-S/N sum. These thresholds yield 4,400~exposures with $t_{\rm exp}=23.5$~days.

When performing the average, individual exposures are weighted by $N_{\ce{H2O}}\,t_{\rm exp}^{1/2}/R_h^2$ to emphasize the highest-activity (i.e., maximum $N_{\ce{H2O}}$), highest-$\mathrm{S/N}$ (i.e., maximum $t_{\rm exp}^{1/2}$), highest $g$-factor (i.e., minimum $R_h^2$) exposures. The resulting sum has a median $\mathrm{S/N} = 200$ in the wavelength region 900-2000~\AA, and an average \ce{H2O} column density of $1.2\times10^{17}~\mathrm{cm}^{-2}$. The large green squares in \autoref{fig:geometry} show the weighted mean value of $R_h$, $\Delta$, $\phi$, and $N_{\ce{H2O}}$ for the grand sum.

\autoref{fig:grandsum} shows the grand sum and our fits to the background (green), CO model (red), and atomic (blue) components. We describe our fitting procedure below.

\subsection{Fitting and Removal of Detected Species}
\label{sec:sum:fits}

\begin{figure}
  \fig{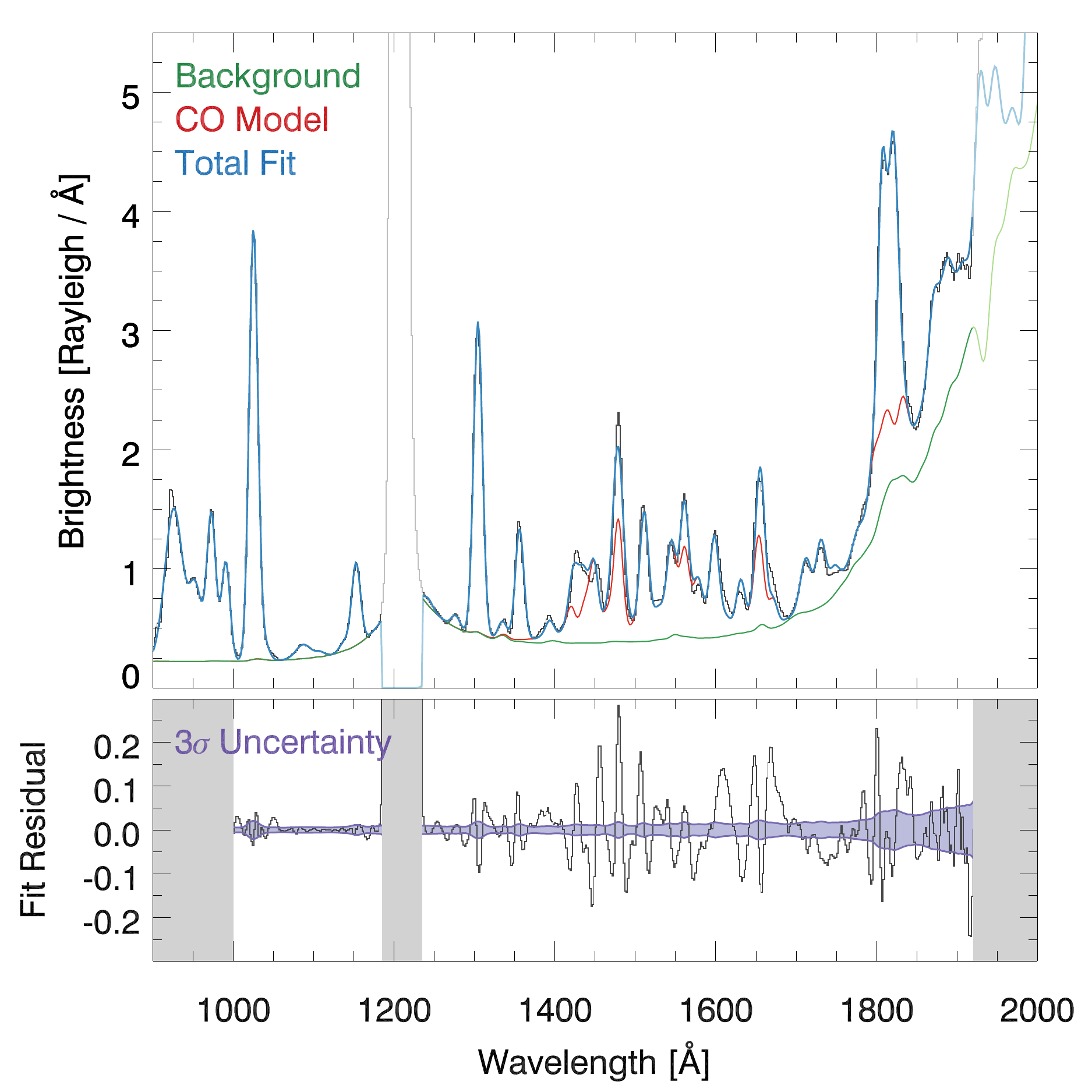}{\columnwidth}{}
  \vspace{-3em}
  \caption{Fit to the ``grand sum'' of Alice exposures near perihelion. \textit{Top:} the grand sum with background (green), CO model (red), and total (blue, includes atomic emissions) fits overlaid. Regions shown in lighter hues are not used to constrain the fits. \textit{Bottom:} the residual of the total fit with $3\sigma$ uncertainty in the grand sum (purple) overlaid. Masked regions (shaded gray) are not used to set upper limits.
  \label{fig:grandsum}}
\end{figure}

As \autoref{fig:spec2d}-\ref{fig:grandsum} demonstrate, Alice detects emissions from several atomic species as well as the CO Fourth Positive and Cameron bands near perihelion. Before we can set limits on undetected species, these ubiquitous emissions must be modeled and removed. To do so, we employ a spectral fitting code first developed for individual Alice exposures \citep{vervack17}.

This code, which will be fully described in a subsequent publication (R. Vervack et~al., 2020, in prep), treats all elements of the spectrum \edit1{physically,} rather than simply fitting the spectrum as a series of Gaussians with an underlying polynomial background or similar approach. \edit1{Specifically, the code deomposes each spectrum} into emission and background components, where the emission represents the signal emitted by the gas species in the coma, and the background represents everything else. This background in turn is composed of several distinct parts: a \edit1{constant} offset or bias, a reflected solar spectrum (from the dust or nucleus), and scattering from H Ly$\alpha$.

The offsets come from light that is scattered internally in the instrument to create uniform backgrounds. The levels differ on either side of Ly$\alpha$ because the detector was coated with different materials: \ce{CsI} redward of Ly$\alpha$ and \ce{KBr} at shorter wavelengths \citep{stern07}. These offsets are a strong function of viewing geometry and cannot be isolated easily; thus, they cannot be removed during pipeline processing.

A reflected solar spectrum is present in nearly all spectra, with the signal being strongest against the sunlit nucleus and varying dramatically in the coma depending on the amount of dust along the line of sight. To account for this reflected solar component, we use publicly available solar spectra \citep[based on SOHO/SUMER between 680-1515~\AA\ and UARS/SOLSTICE above 1515~\AA;][]{wilhelm97,rottman93} for each date during the escort phase, scaled to the heliocentric position of 67P/C-G, and convolved to match the spectral resolution of Alice. For individual exposures, these daily solar spectra are interpolated to the appropriate time of day, but when fitting our sum we use the daily spectrum from the weighted mean date (\autoref{fig:geometry}).

Three factors are then applied in the fitting to match the reflected solar component in the observed spectrum. First, the overall solar flux is scaled down to account for the signal reduction upon reflection. This can be due to a number of things: the slit only partially sampling the dayside nucleus and coma, albedo effects, the amount of dust along the line of sight, and so on. Second, the solar spectrum is ``reddened'' by applying a linear function that tilts the spectrum as a function of wavelength.

The third factor is that the solar spectrum is adjusted to account for the presence of \ce{H2O} vapor, which absorbs the sunlight as a function of wavelength. All sunlight reaching the instrument must pass through the coma and thus experiences such absorption. However, the line of sight experiencing this absorption (Sun to comet, then comet to \rosetta) differs from the line of sight seen by Alice. Thus, we adopt the $N_{\ce{H2O}}$ predictions of the modified \citet{hansen16} model described in \autoref{sec:sum} for the Alice line of sight, which \citet{keeney19} showed are consistent with Alice absorption measurements against the continuum of background stars, rather than the values produced as part of the spectral fitting.

The final background component is the scattered light from Ly$\alpha$. This is treated as two different exponential functions on either side of the line; however, care must be employed in the fitting of these exponentials. Significant degradation of the Ly$\alpha$ region occurred on the detector over the escort phase \citep{pineau19}, leading to changes in the observed spectral shape and absolute flux of the line that are difficult to quantify. To account for this, we do not fit the core of the Ly$\alpha$ line but only fit the wings of the line away from the degraded region.

We fit the CO Fourth Positive and Cameron bands with model spectra following \citet{lupu07} and \citet{conway81}, respectively. The atomic lines are treated as Gaussians, which closely mimic the actual line shape but provide for an analytical treatment. All of the various components of the spectrum are fit simultaneously, with the red (longer wavelengths) and blue (shorter wavelengths) sides of Ly$\alpha$ treated separately. Simultaneously fitting the components is key in resolving the spectral confusion at wavelengths where multiple emissions and background features overlap.

\autoref{fig:grandsum} shows the results of applying this code to the grand sum of Alice perihelion exposures. The top panel displays the sum itself in black, the best-fit background in green, the best-fit CO Fourth Positive and Cameron band emission in red, and the total fit including atomic emissions in blue. The bottom panel shows the fit residuals, which are used to set upper limits in \autoref{sec:limits}, compared to the $3\sigma$ uncertainty in the coaddition.

For most of the fitted wavelength range, the amplitude of the fit residuals is larger than the $3\sigma$ uncertainty in the grand sum. There are several contributing factors. First, because we average rows with different spectral resolutions when creating our sum, the resulting line spread function is not purely Gaussian. Second, the solar spectrum used in the background fit is from a single day but the spectrum being fit is an average. Finally, the CO model spectra assume purely fluorescent emission \citep{lupu07,conway81}, except for a broad pseudocontinuum centered at $\sim1565$~\AA\ that underlies the CO Fourth Positive band and is characteristic of excitation by low-energy electrons \citep{ajello19}.

We believe this last factor is the most important because electron impact excitation affects the atomic emission lines \citep{feldman15,feldman16} and CO Fourth Positive band \citep{feldman18} in Alice data further from perihelion. Indeed, the regions of the spectrum with the largest residuals are associated with mismatches between the measured and modeled CO emission. Nevertheless, the discrepancies between the grand sum and the best-fit model, although large compared to the uncertainty in the high-S/N coaddition, are small enough that sensitive upper limits can still be achieved.

\subsection{Rare Lines Search}
\label{sec:sum:rare}

We have also searched the individual exposures that contribute to the grand sum (\autoref{fig:geometry}) for instances of strong, rare lines that are not attributable to ubiquitous H, O, C, S, or CO emissions. This search was performed with two independent spectral fitting codes: the first is the code described in \autoref{sec:sum:fits}, and the second is a Python routine that fits emission lines with skewed Gaussians atop a piece-wise continuous background \citep{noonan18}. Neither fitting code found any cases where previously unidentified features were detected. This analysis suggests that averaging many exposures together to increase S/N is not diluting the presence of otherwise detectable features present in a small number of low-S/N individual exposures.

\section{Setting Upper Limits}
\label{sec:limits}

Due to the complex structure of the fit residuals in the bottom panel of \autoref{fig:grandsum}, care must be taken to set robust limits. We employ a bootstrapping procedure to directly integrate the residuals, modified by a randomly drawn amount from the uncertainty distribution as a function of wavelength. The integration is performed over a width of 12.9 and 12.6~\AA, respectively, blueward and redward of Ly$\alpha$ (i.e., the filled-slit resolution of the sum). The exercise is repeated 10,000~times for each line, allowing the center of the integration window to vary slightly from iteration to iteration to account for uncertainties in the absolute wavelength calibration. Finally, we analyze the integrated brightness values near each wavelength of interest, selecting the 99th-percentile value as the brightness limit for that line  (i.e., 1\% of the iterations found a larger brightness than the adopted value). These brightness limits are shown in the fourth column of \autoref{tab:lims}.

Many of the species in \autoref{tab:lims} have multiple transitions in the Alice bandpass, but only the transition that results in the lowest column density limit, $N_{\rm lim}$, is listed. The column density and brightness are related by $B=gN$, so the most stringent $N_{\rm lim}$ can be obtained by minimizing $B_{\rm lim}$ or maximizing the fluorescence efficiency factor, $g$. Furthermore, only the wavelength ranges 1000-1185~\AA\ and 1235-1920~\AA\ are considered due to the influence of the instrumental artifact at wavelengths $\lesssim1000$~\AA\ \citep{noonan16}, the unmodeled core of Ly$\alpha$ from 1185-1235~\AA, and the declining Alice sensitivity at wavelengths $\gtrsim1900$~\AA\ \citep{stern07}. This procedure assumes that unobserved emissions are produced by solar resonance fluorescence, despite the fact that most of the observed emissions can be produced partially or wholly by electron impact excitation.

\begin{deluxetable*}{lCCCCl}

\tablecaption{Alice Upper Limits near Perihelion
\label{tab:lims}}

\tablehead{
  \colhead{Species} &
  \colhead{Wavelength} &
  \colhead{$\langle g \rangle$} &
  \colhead{$B_{\rm lim}$} &
  \colhead{$N_{\rm lim}$} &
  \colhead{Atomic Data Reference} \\
  \colhead{} &
  \colhead{(\AA)} &
  \colhead{($\mathrm{ph\,s^{-1}}$)} &
  \colhead{(R)} &
  \colhead{($\mathrm{cm^{-2}}$)} &
  \colhead{}
  }

\colnumbers
\startdata
Al      & 1766.1 & 1.22\times10^{-4} & <  -0.13 &           \nodata & \citet{vanhoof18} \\
Ar      & 1048.2 & 7.30\times10^{-8} & <   0.21 & <2.8\times10^{12} & \citet{vanhoof18} \\
As      & 1890.4 & 4.77\times10^{-4} & <   0.77 & <1.6\times10^{ 9} & \citet{kurucz99}  \\
Au      & 1879.8 & 1.20\times10^{-5} & <   0.64 & <5.3\times10^{10} & \citet{kurucz99}  \\
B       & 1826.2 & 5.72\times10^{-5} & <    1.9 & <3.3\times10^{10} & \citet{vanhoof18} \\
Ca      & 1883.2 & 8.91\times10^{-6} & <   0.48 & <5.4\times10^{10} & \citet{kurucz99}  \\
Cl      & 1335.7 & 5.99\times10^{-6} & <   0.13 & <2.1\times10^{10} & \citet{vanhoof18} \\
Co      & 1850.3 & 9.07\times10^{-6} & <   0.77 & <8.4\times10^{10} & \citet{vanhoof18} \\
Fe      & 1851.7 & 5.72\times10^{-6} & <   0.55 & <9.7\times10^{10} & \citet{vanhoof18} \\
Hg      & 1849.5 & 3.45\times10^{-4} & <   0.91 & <2.6\times10^{ 9} & \citet{kramida18} \\
Kr      & 1235.8 & 1.74\times10^{-7} & <    29. & <1.7\times10^{14} & \citet{cashman17} \\
Mg      & 1827.9 & 9.08\times10^{-5} & <    2.2 & <2.5\times10^{10} & \citet{vanhoof18} \\
Mn      & 1785.5 & 5.98\times10^{-6} & <   0.79 & <1.3\times10^{11} & \citet{kurucz99}  \\
N       & 1134.6 & 1.42\times10^{-7} & <  0.031 & <2.2\times10^{11} & \citet{vanhoof18} \\
N$^+$   & 1085.1 & 4.67\times10^{-7} & <  0.028 & <6.0\times10^{10} & \citet{vanhoof18} \\
P       & 1774.9 & 2.64\times10^{-5} & <   0.43 & <1.6\times10^{10} & \citet{vanhoof18} \\
Sc      & 1742.4 & 1.10\times10^{-5} & <   -1.2 &          \nodata  & \citet{kurucz99}  \\
Si      & 1848.9 & 7.22\times10^{-5} & <   0.95 & <1.3\times10^{10} & \citet{vanhoof18} \\
Xe      & 1469.6 & 9.50\times10^{-7} & <    1.3 & <1.4\times10^{12} & \citet{kramida18} \\
Zn      & 1589.6 & 2.49\times10^{-6} & <   0.59 & <2.4\times10^{11} & \citet{vanhoof18} 
\enddata

\tablecomments{Limits are quoted at the 99\% confidence level.}

\end{deluxetable*}

The third column of \autoref{tab:lims} shows the $g$-factor adopted for each species. We calculated $g$-factors for each exposure that contributes to the grand sum using atomic line data (e.g., oscillator strengths, Einstein $A$ values) for each transition as referenced in the final column of \autoref{tab:lims}. When necessary, a temperature of 100~K is assumed, and we utilize the same daily solar spectrum used to fit individual exposures (\autoref{sec:sum:fits}), but at native resolution (i.e., not convolved to the Alice spectral resolution) so we can accurately account for the Doppler shift introduced by the comet's relative velocity to the Sun \citep{swings41} at the time of observation. The tabulated value, $\langle g \rangle$, is the weighted mean of the exposure-level $g$-factors, where each exposure has the same weight ($N_{\ce{H2O}}\,t_{\rm exp}^{1/2}/R_h^2$) used when creating the sum. The fifth column of \autoref{tab:lims} shows the column density limit for each species. \edit1{We do not list column density limits for Al or Sc because their brightness limits are} \edit2{unphysical (i.e., negative) due to systematically negative residuals between wavelengths of $\sim1720$ and 1750~\AA\ (see bottom panel of \autoref{fig:grandsum}); thus, we cannot set physically meaningful column density limits on these species.}

\section{Discussion}
\label{sec:disc}

\subsection{Comparisons with ROSINA Measurements}
\label{sec:disc:rosina}

Although ROSINA/DFMS primarily cataloged the molecular constituents of 67P/C-G \citep[e.g.,][]{leroy15,fougere16,altwegg17,hoang17,lauter19}, the abundances of several atomic species have also been reported. \autoref{tab:rosina} lists the relative abundance $N_{\rm lim}/N_{\ce{H2O}}$ near perihelion for comparison with ROSINA measurements at other times in the escort mission, where $N_{\ce{H2O}}$ was determined as described in \autoref{sec:sum}.

\begin{deluxetable*}{lCCCl}

\tablecaption{Comparison with ROSINA/DFMS Measurements
\label{tab:rosina}}

\tablehead{
  \colhead{Species} &
  \colhead{$N_{\rm lim}/N_{\ce{H2O}}$} &
  \colhead{$({\rm X}/\ce{H2O})_{\rm DFMS}$} &
  \colhead{Alice\,/\,DFMS} &
  \colhead{DFMS Reference}
  }

\colnumbers
\startdata
Ar      & < 2.3\times10^{-5} & (5.8\pm2.2)\times10^{-6} & <    4.0 & \citet{rubin18} \\
Ca      & < 4.5\times10^{-7} & (1.1\pm0.8)\times10^{-6} & <   0.41 & \citet{wurz15}  \\
Kr      & < 1.4\times10^{-3} & (4.9\pm2.2)\times10^{-7} & <   2900 & \citet{rubin18} \\
Si      & < 1.1\times10^{-7} & (7.3\pm5.8)\times10^{-5} & < 0.0015 & \citet{wurz15}  \\
Xe      & < 1.2\times10^{-5} & (2.4\pm1.1)\times10^{-7} & <    50. & \citet{rubin18}
\enddata

\tablecomments{Limits are quoted at the 99\% confidence level. The column density of water for the grand sum is $\langle N_{\ce{H2O}} \rangle = 1.2\times10^{17}~\mathrm{cm^{-2}}$.}

\end{deluxetable*}

Quantifying the noble gas abundance of 67P/C-G is of particular interest because noble gases are both chemically inert and extremely volatile, making them excellent tracers of a comet's thermal evolution \citep{stern99}. ROSINA has detected Ar \citep{balsiger15}, Kr \citep{rubin18}, and Xe \citep{marty17} in the coma of 67P/C-G, allowing us to assess the sensitivity of our upper limits compared to their detections. ROSINA did not detect Ne, but set an upper limit of $\ce{Ne}/\ce{H2O} < 5\times10^{-8}$ \citep{rubin18}; unfortunately, the Ne doublet (736, 744~\AA) is heavily affected by an instrumental artifact in Alice data \citep{noonan16} so we are not able to set an independent limit.

ROSINA/DFMS detected \ce{^{36}Ar} and \ce{^{38}Ar} from 2014~October 19-23 ($R_h=3.1$~AU) when \rosetta\ was executing very close fly-bys $\sim10$~km from the comet center \citep{balsiger15}. The measured isotopic ratio $\ce{^{36}Ar}/\ce{^{38}Ar} = 5.4\pm1.4$ was similar to values for the Earth and the solar wind \citep{balsiger15}, and the abundance relative to \ce{H2O} was found to be $\ce{^{36}Ar}/\ce{H2O} = (0.1$-$2.3)\times10^{-5}$. When \rosetta\ was once again executing very close fly-bys of 67P/C-G, this time long after perihelion (2016~May 14-31; $R_h=3.0$-3.1~AU), ROSINA/DFMS detected 7~isotopes\footnote{The detected isotopes are \ce{^{128}Xe}, \ce{^{129}Xe}, \ce{^{130}Xe}, \ce{^{131}Xe}, \ce{^{132}Xe}, \ce{^{134}Xe}, and \ce{^{136}Xe}.} of Xe \citep{marty17}. Unlike Ar, the Xe isotopes were not consistent with terrestrial or solar wind values \citep[see Figure~1 of][]{marty17}. Finally, \citet{rubin18} reported that during the same time period that Xe was measured, five isotopes\footnote{The detected isotopes are \ce{^{80}Kr}, \ce{^{82}Kr}, \ce{^{83}Kr}, \ce{^{84}Kr}, and \ce{^{86}Kr}.} of Kr were also detected, with slight deviations relative to solar wind values.

Furthermore, \citet{rubin18} estimated the bulk abundances of the noble gases in 67P/C-G near perihelion, which was complicated by the fact that the noble gases were only detected during close fly-bys well away from perihelion. Nonetheless, the noble gas abundances near perihelion were estimated by taking advantage of the facts that Ar and \ce{N2} are highly correlated in 67P/C-G \citep{balsiger15,rubin18}, and that the relative $\ce{N2}/\ce{H2O}$ abundance was measured by DFMS throughout \rosetta's escort mission. The resulting estimates of the bulk abundance of Ar, Kr, and Xe with respect to \ce{H2O} \citep{rubin18} are shown in the third column of \autoref{tab:rosina}.

The fourth column of \autoref{tab:rosina} shows the ratio of the Alice upper limit to the DFMS measurement. All of the Alice noble gas limits are larger than, and therefore consistent with, the ROSINA values. The Alice Ar and Xe limits are a factor of 4 and 50 larger, respectively, than the bulk abundances from \citet{rubin18}, but the Alice Kr limit is much higher due to blending of the Kr 1236~\AA\ line with the wing of Ly$\alpha$.

During the same close fly-by when Ar was first detected in 67P/C-G, ROSINA also detected atomic Na, K, Si, and Ca, presumably sputtering off dust grains on the surface \citep{wurz15}. Although Alice is not sensitive to neutral Na or K, limits for Si and Ca were placed (\autoref{tab:lims}). For various technical reasons, some other species in \autoref{tab:lims} that may also be expected to sputter off cometary dust grains (e.g., Mg, Fe, Al) cannot be measured by ROSINA \citep{wurz15}.

The relative abundance $\ce{Si}/\ce{H2O}$ in \autoref{tab:rosina} was estimated from Figure~3 of \citet{wurz15}, and $\ce{Ca}/\ce{H2O}$ was derived from the estimated $\ce{Si}/\ce{H2O}$ value using the Ca/Si ratios in Table~2 of \citet{wurz15}. \citet{wurz15} found that the $\ce{Si}/\ce{H2O}$ abundance varied over a wide range because the Si was detected primarily over the southern hemisphere, where the solar wind was able to access the surface, whereas \ce{H2O} was released primarily from the northern hemisphere at that time. We note that \citet{rubin17} also reported on the Si abundance of 67P/C-G, finding that the heavy Si isotopes are depleted, but since they measured ionic abundances (e.g., \ce{$^{28}$Si$^+$}, \ce{$^{29}$Si$^+$}, \ce{$^{30}$Si$^+$}) we restrict our comparisons to the neutral Si abundances of \citet{wurz15}.

Unlike our noble gas limits, the Alice limits on Si and Ca are lower than the values measured by DFMS. However, this is not surprising since \citet{wurz15} interpreted their detections as the result of sputtering of cometary dust grains by solar wind ions. Sputtering is not expected near perihelion (i.e., the time frame over which the Alice limits are valid) because the \edit2{solar wind cavity \citep{behar17}} and diamagnetic cavity \citep{goetz16} prevent the solar wind from having direct access to the comet's surface, and Si and Ca from being sputtered into the coma.

\newpage
\begin{deluxetable}{lCCC}

\tablecaption{Comparison with Meteoritic and Solar Abundances
\label{tab:solar}}

\tablehead{
  \colhead{Species} &
  \colhead{$\log{(N_{\rm lim}/N_{\rm O})}$} &
  \colhead{$\log{({\rm X/O})_{\rm CI}}$} &
  \colhead{$\log{({\rm X/O})_{\Sun}}$}
  }

\colnumbers
\startdata
Ar    & < +0.6 & -8.90        & -2.29\pm0.14 \\
As    & < -2.6 & -6.10\pm0.06 &      \nodata \\
Au    & < -1.1 & -7.60\pm0.06 & -7.77\pm0.11 \\
B     & < -1.3 & -5.61\pm0.06 & -5.99\pm0.21 \\
Ca    & < -1.1 & -2.11\pm0.04 & -2.35\pm0.06 \\
Cl    & < -1.5 & -3.17\pm0.07 & -3.19\pm0.30 \\
Co    & < -0.9 & -3.53\pm0.04 & -3.70\pm0.09 \\
Fe    & < -0.9 & -0.95\pm0.04 & -1.19\pm0.06 \\
Hg    & < -2.4 & -7.23\pm0.09 &      \nodata \\
Kr    & < +2.4 & -10.67       & -5.44\pm0.08 \\
Mg    & < -1.5 & -0.87\pm0.04 & -1.09\pm0.06 \\
Mn    & < -0.7 & -2.92\pm0.04 & -3.26\pm0.06 \\
N     & < -0.5 & -2.14\pm0.07 & -0.86\pm0.07 \\
N$^+$ & < -1.1 &      \nodata &      \nodata \\
P     & < -1.6 & -2.97\pm0.06 & -3.28\pm0.06 \\
Si    & < -1.7 & -0.89\pm0.04 & -1.18\pm0.06 \\
Xe    & < +0.3 & -10.35       & -6.45\pm0.08 \\
Zn    & < -0.5 & -3.77\pm0.06 & -4.13\pm0.07
\enddata

\tablecomments{Limits are quoted at the 99\% confidence level. The column density of oxygen for the grand sum is $\langle N_{\rm O} \rangle = 7.0\times10^{11}~\mathrm{cm^{-2}}$. Abundance ratios for CI carbonaceous chondrites and the solar photosphere are from \citet*{lodders09} and \citet{asplund09}, respectively.}

\end{deluxetable}

\subsection{Comparisons with Meteoritic and Solar Abundances}
\label{sec:disc:solar}

We compare our column density limits to CI carbonaceous chondrite \citep{lodders09} and solar photospheric \citep{asplund09} abundances in \autoref{tab:solar}. We choose to compare to CI carbonaceous chondrite abundances because they are well-known and consistent with the mean elemental abundances in {\sl Stardust} samples of Comet 81P/Wild~2 \citep{flynn06}. \edit1{Note that \citet{asplund09} use indirect photospheric estimates for the noble gases.}

To facilitate these comparisons, we normalize our column density limits by the column density of O, which we derive from the best-fit brightness to the O 1302~\AA\ line using $\langle g_{\rm O} \rangle = 1.70\times10^{-5}~\mathrm{photons\,s^{-1}}$. We normalize by O instead of H because it is less volatile and has no contamination from the interplanetary medium \citep[e.g.,][]{vincent14}. For all but the most volatile elements (e.g., N \edit1{and the noble gases}), the meteoritic and photospheric abundances relative to O are comparable, and often indistinguishable within the statistical uncertainties.

The Alice upper limits are orders of magnitude above the expected ratios, except for Fe/O, Mg/O, \edit1{and Si/O}. Fe, Mg, \edit1{and Si are all} refractory elements common to mineral grains, and only expected to enter the coma via sputtering. \citet{wurz15} suggest that the amounts of Mg and Fe sputtered may be comparable to the amount of Si, but ROSINA/DFMS cannot constrain these species due to low detection efficiencies and interferences. Thus, we attribute the low Fe/O, Mg/O, \edit1{and Si/O} limits found by Alice near perihelion to the presence of the comet's \edit2{solar-wind and} diamagnetic cavities.

\section{Summary}
\label{sec:summary}

We have presented upper limits for 18~species in the coma of 67P/C-G, derived from a ``grand sum'' of exposures obtained by \rosetta's Alice far-UV imaging spectrograph near perihelion. Essential steps in our procedure were carefully selecting and vetting which exposures contribute to the coaddition, and fitting and removing backgrounds from reflected solar continuum and gaseous \ce{H2O} vapor in the coma and ubiquitous emissions from H, O, C, S, and CO. Upper limits were derived by direct integration of the fit residuals.

Our upper limits for the noble gases Ar and Xe were a factor of 4 and 50, respectively, larger than estimates of the bulk abundance relative to \ce{H2O} derived from ROSINA measurements \citep{rubin18}. On the other hand, our upper limits for Ca and Si were a factor of 2.4 and 670 below those observed by ROSINA in 2014 October \citep{wurz15}, offering indirect evidence that the \edit2{solar-wind and} diamagnetic cavities of 67P/C-G \citep{behar17,goetz16} prevented the solar wind from sputtering dust grains on the nucleus near perihelion. Our upper limits on Fe/O and Mg/O compared to meteoritic \citep{lodders09} and solar \citep{asplund09} abundances also support this conclusion.

\acknowledgments
\rosetta\ is an ESA mission with contributions from its member states and NASA. We thank the members of the \rosetta\ Science Ground System and Mission Operations Center teams, in particular Richard Moissl and Michael K\"uppers, for their expert help in planning and executing the Alice observations. The Alice team acknowledges support from NASA via Jet Propulsion Laboratory contract 1336850 to the Southwest Research Institute.

\facility{\rosetta\ (Alice)}
\software{NAIF/SPICE \citep{acton96}}

\bibliographystyle{aasjournal}
\bibliography{references}

\end{document}